\documentclass[a4paper, amsfonts, amssymb, amsmath, reprint, showkeys, nofootinbib, twoside]{revtex4-1}
\usepackage[english]{babel}
\usepackage[utf8]{inputenc}
\usepackage[colorinlistoftodos, color=green!40, prependcaption]{todonotes}
\usepackage{amsthm}
\usepackage{ulem}
\usepackage{multirow} 
\usepackage{xcolor}
\usepackage{graphicx}
\usepackage{caption}
\usepackage[T1]{fontenc}
\usepackage[pdftex, pdftitle={Article}, pdfauthor={Author}]{hyperref} 
\usepackage{booktabs}
\usepackage{siunitx}
\definecolor{purple}{rgb}{0.58,0.0,0.83}

\definecolor{blue(pigment)}{rgb}{0.2, 0.2, 0.6}

\definecolor{red}{rgb}{1.0, 0.0, 0.0}

\usepackage{scalerel}
\usepackage{tikz}
\usetikzlibrary{svg.path}

\definecolor{orcidlogocol}{HTML}{A6CE39}
\tikzset{
  orcidlogo/.pic={
    \fill[orcidlogocol] svg{M256,128c0,70.7-57.3,128-128,128C57.3,256,0,198.7,0,128C0,57.3,57.3,0,128,0C198.7,0,256,57.3,256,128z};
    \fill[white] svg{M86.3,186.2H70.9V79.1h15.4v48.4V186.2z}
                 svg{M108.9,79.1h41.6c39.6,0,57,28.3,57,53.6c0,27.5-21.5,53.6-56.8,53.6h-41.8V79.1z M124.3,172.4h24.5c34.9,0,42.9-26.5,42.9-39.7c0-21.5-13.7-39.7-43.7-39.7h-23.7V172.4z}
                 svg{M88.7,56.8c0,5.5-4.5,10.1-10.1,10.1c-5.6,0-10.1-4.6-10.1-10.1c0-5.6,4.5-10.1,10.1-10.1C84.2,46.7,88.7,51.3,88.7,56.8z};
  }
}

\newcommand\orcidicon[1]{\href{https://orcid.org/#1}{\mbox{\scalerel*{
\begin{tikzpicture}[yscale=-1,transform shape]
\pic{orcidlogo};
\end{tikzpicture}
}{|}}}}

\begin{document}

\title{Cosmological evolution driven by polytropic fluids in an inhomogeneous spacetime}

\author{Gilberto Aguilar-Pérez$^{a}$\orcidicon{0000-0001-6821-4564}}
\email{gilaguilar@uv.mx}

\author{Miguel Cruz$^{a}$\orcidicon{0000-0003-3826-1321}}
\email{miguelcruz02@uv.mx}

\author{Mohsen Fathi$^{b}$\orcidicon{0000-0002-1602-0722}}
\email{mohsen.fathi@ucentral.cl}

\author{Daniel de Jesús García-Castro$^{a}$}
\email{zS20015181@estudiantes.uv.mx}

\author{J. R. Villanueva$^{c}$\orcidicon{0000-0002-6726-492X}}
\email{jose.villanueva@uv.cl}

\affiliation{$^{a}$ Facultad de F\'{\i}sica, Universidad Veracruzana 91097, Xalapa, Veracruz, M\'exico\\
$^{b}$ Centro de Investigaci\'{o}n en Ciencias del Espacio y F\'{i}sica Te\'{o}rica, Universidad Central de Chile, La Serena 1710164, Chile\\
$^{c}$ Instituto de Física y Astronomía, Universidad de Valparaíso, Gran Bretaña 1111, Valparaíso, Chile}

\begin{abstract}
Addressing the late-time accelerated expansion of the universe, known as the ``dark energy problem'', remains a central challenge in cosmology. While the cosmological constant is the standard explanation, alternative models such as quintessence, phantom fluids, and Chaplygin gas have been proposed. This work investigates the generalized Chaplygin gas (GCG) model, which is characterized by a polytropic equation of state. We explore this model within the framework of an anisotropic fluid, by means of a metric that reduces to the standard form of the Friedmann-Lema\^itre-Robertson-Walker (FLRW) spacetime at cosmological scales. To assess the model's viability, we derive analytical expressions for the scale factor, the Hubble parameter, and the deceleration parameter. Finally, the model is tested against observational data to constrain its parameters and evaluate its consistency.   
\end{abstract}

\maketitle

\section{Introduction}
\label{sec:intro}

Cosmology, the study of the universe as a whole, has been successfully guided by the cosmological principle, which establishes that the universe is homogeneous and isotropic on large scales \cite{1, Weinberg:2008zzc, 2, peebles1993principles}. This principle is a cornerstone of the standard model of cosmology, known as $\Lambda$CDM, which is based on the FLRW metric. The $\Lambda$CDM model impressively explains a vast range of observations, from the anisotropies in the Cosmic Microwave Background (CMB) \cite{2020} to the distribution of large-scale structures. However, the model relies on two enigmatic components: dark matter and dark energy, the latter in the form of a cosmological constant, dubbed as $\Lambda$. The cosmological constant is plagued by severe theoretical issues, such as the fine-tuning problem—a huge discrepancy between the observed value and theoretical predictions \cite{Carroll_2001}—and the cosmic coincidence problem \cite{Velten_2014}. These foundational challenges, coupled with growing observational tensions like the Hubble constant discrepancy \cite{Riess_2016} and evidence of inhomogeneities on various scales, provide strong motivation for exploring alternative cosmological models.\\

One prominent avenue of research involves relaxing the assumption of perfect homogeneity. Several exact solutions to Einstein's equations that describe inhomogeneous universes have been proposed, most notably the Lema\^itre-Tolman-Bondi (LTB) and Szekeres models, see for instance \cite{Modan_2024, Koksbang_2022, Koksbang_2021, Koksbang_2019, f045d741-d862-34d0-962c-d170399d61e6, Szekeres:1974ct, bolejko2017inhomogeneous, bolejko2011inhomogeneous} and references therein. The LTB model, which describes a spherically symmetric but radially inhomogeneous dust-filled universe, has been widely explored in the literature. It has been used to test the hypothesis that the observed cosmic acceleration is not due to a new energy component (such as dark energy) but is instead an apparent effect arising from our location within a large-scale underdense region, or void \cite{Buchert_2007}. While such models face challenges in simultaneously fitting all cosmological data, they demonstrate that inhomogeneity can significantly impact the interpretation of cosmic expansion.\\

More recent approaches combine inhomogeneity with more complex fluid descriptions. For instance, models incorporating an anisotropic fluid within an inhomogeneous metric have been shown to generate an effective dark energy component from the interaction between geometry and matter, see Refs. \cite{cadoni2018effective,cadoni2020anisotropic}. Following this line of reasoning, this paper investigates the cosmological dynamics driven by a fluid with a generalized Chaplygin gas (GCG) equation of state, a specific type of polytropic fluid. The GCG is particularly compelling as it can smoothly interpolate between a pressureless matter-dominated phase at early times and a negative-pressure, dark-energy-dominated phase at late times, thus offering a unified description of the dark sector \cite{kamenshchik2001alternative,delCampo:2009cz}. Our work explores whether such a unified fluid model, when embedded within an inhomogeneous and anisotropic spacetime, can provide a viable and self-consistent description of the observed universe, such as the late-time accelerated expansion.\\

The structure of this work is as follows: In Section \ref{sec:sec}, a spacetime characterizing inhomogeneities is studied. However, such spacetime reduces to the typical FLRW in the cosmological limit. Additionally, the energy-momentum tensor of an anisotropic fluid is introduced to derive cosmological equations that incorporate the effects of inhomogeneities in the universe. Section \ref{sec:chaply} aims to present two types of Chaplygin models. The main characteristic of these models is that they introduce a polytropic equation of state. The cosmological consequences of Chaplygin-type models are analyzed. In Section \ref{sec:disscu}, we present the results of introducing Chaplygin-type equations of state into the equations describing a universe with inhomogeneities, with the aim of studying the evolution of the universe from a more realistic perspective. Finally, Section \ref{sec:obs} is devoted to testing the viability of the model based on cosmological observations. 
This research has been carried out using units in which $8\pi G = c = 1$.
 
\section{Geometry and matter content}
\label{sec:sec}

We will consider the evolution of a universe characterized by an inhomogeneous space-time and an energy-momentum tensor of an anisotropic fluid; we implement the procedure developed by Cadoni, Sanna 
and Tuveri in Ref. \cite{cadoni2020anisotropic}. The spacetime is characterized by the following Schwarzschild-like metric
\begin{equation}\label{1}
    ds^2=a^2(\eta)\left[-f(r)e^{\gamma(r)}d\eta^2+\frac{dr^2}{f(r)}+r^2d\Omega^2\right],
\end{equation}
where $d\Omega^2=d\theta^2+\sin^2\theta d\phi^2$.
The metric functions $f(r)$, $\gamma(r)$
depend only on the radial coordinate $r$,
while the function $a(\eta)$ depends on the conformal time $\eta$. 
The energy-momentum tensor is described by the following expression 
\begin{equation}\label{2}
    T_{\mu\nu}=(\rho+p_t)u_\mu u_\nu+p_tg_{\mu\nu}-(p_t-p_r)\omega_\mu\omega_\nu,
\end{equation}
where $u_\mu$ and $\omega_\nu$ label vectors that 
satisfy the condition $u^\nu u_\nu=-1$, $\omega^\nu \omega_\nu=1$ and $u^\mu\omega_\mu=0$, 
while $p_t$ and $p_r$ are the tangential and radial components of the pressure, respectively. 
As can be seen, the two different components of the pressure give rise to anisotropies. For a general treatment of the energy-momentum tensor see Refs. \cite{cosenza1981some,herrera1997local}. 
Of course, with $p_r=p_t$ the anisotropy term 
vanishes and an energy-momentum tensor for a perfect fluid is recovered. 
One way to guarantee that $p_r\neq p_t$, that is, a nontrivial anisotropy, is that $p_r$, $p_t$ and $\rho$ depend on $r$ and on $\eta$. In doing so, we also allow for a description of a universe in three different regimes: Newtonian, galactic, and cosmological; it may happen that when $r\rightarrow \infty$, $p_r=p_t$, so we can recover a FLRW-like behavior. 
Assuming an adequate frame, 
the Einstein equations read as follows
\begin{equation}\label{3}
    3\left(\frac{\dot{a}}{a}\right)^2-\frac{e^\gamma f}{r^2}\left(1-f+rf^\prime\right)=a^2\rho fe^\gamma,
\end{equation}
\begin{equation}\label{4}
    \frac{\dot{a}}{af}\left(f^\prime+f\gamma^\prime\right)=0,
\end{equation}
\begin{small}
\begin{equation}\label{5}
    \frac{e^{-\gamma}}{(raf)^2}\left[r^2\dot{a}^2+e^\gamma a^2f(-1+f+rf^\prime+rf\gamma^\prime)-2r^2a\ddot{a}\right]=p_r\frac{a^2}{f}.
\end{equation}
\end{small} 
Here, the dots and primes indicate derivatives with respect to $\eta$ and $r$, respectively. 
Eq. \eqref{4} leads to two cases: $\dot{a}=0$ and $f^\prime+f\gamma^\prime=0$. The first case corresponds to a static universe and is not of our interest. The second case relates the two $r$-dependent metric functions
\begin{equation}\label{8}
    f^\prime=-f\gamma^\prime.
\end{equation}
From the condition of conservation of the energy-momentum tensor $\nabla_\mu T^{\mu\nu}=0$, emerge two additional equations given by
\begin{equation}\label{6}
    \dot{\rho}+\frac{\dot{a}}{a}\left(3\rho+p_r+2p_t\right)=0,
\end{equation}
and
\begin{equation}\label{7}
    p_r^\prime+\frac{2}{r}\left(p_r-p_t\right)+\frac{1}{2}\left(p_r+\rho\right)\left(\gamma^\prime+\frac{f^\prime}{f}\right)=0.
\end{equation}
Now plugging \eqref{8} into \eqref{7} allows us to write
\begin{equation}\label{9}
    p_t=p_r+\frac{r}{2}p_r^\prime,
\end{equation}
and from \eqref{8} we can explicitly solve $f$ in terms of $\gamma$, yielding 
\begin{equation}\label{10}
    f=e^{-\gamma}.
\end{equation}
Using Eqs. \eqref{9} and \eqref{10}, thus the Eqs. \eqref{3}, \eqref{5} and \eqref{6} can be rewritten in the following form 
\begin{equation}\label{11}
    3\left(\frac{\dot{a}}{a}\right)^2+\frac{1-f-rf^\prime}{r^2}=a^2\rho,
\end{equation}
\begin{equation}\label{12}
    \left(\frac{\dot{a}}{a}\right)^2-2\frac{\ddot{a}}{a}-a^2p_r=\frac{1-f}{r^2},
\end{equation}
\begin{equation}\label{13}
    \dot{\rho}+\frac{\dot{a}}{a}\left[3\rho+3p_r+rp_r^\prime\right]=0.
\end{equation}
We now focus on finding the functional form of $f(r)$. The first term of the left-hand side of Eq. \eqref{11} depends only on $\eta$ and the second term depends only on $r$, in this way $3\mathcal{H}(\eta)+\mathcal{E}(r)=a^2\rho(\eta,r)$, where we have defined
\begin{equation}\label{14}
    \mathcal{E}(r):=\frac{1-f-rf^\prime}{r^2},
\end{equation}
rearranging this expression, we note that it can be written in the following form 
\begin{equation}\label{15}
    1-r^2\mathcal{E}(r)=(rf)^\prime,
\end{equation}
which can be integrated with respect to $r$: 
\begin{equation}\label{16}
    f(r)=1-\frac{m_B(r)}{r}-\frac{M}{r},
\end{equation}
with $m_B(r):=\int r^2\mathcal{E}(r)\;dr$ and $M$ an integration constant, in fact $m_B$ is the Misner-Sharp mass \cite{cadoni2020anisotropic}. For a spacetime with $k=0$, we can set $m_B(r)=\frac{c_1}{r}$, where $c_1$ is a constant with appropriate units. In doing so, $f(r)\rightarrow 1$ and consequently, the spacetime becomes the usual FLRW; we call this the cosmological limit. 

According to the conditions found on the left-hand side of Eq. \eqref{11}, we can assume $a^2(\eta)\rho(\eta,r) := a^2(\eta)\tilde{\rho}(\eta) + \mathcal{E}(r)$, leading to 
\begin{equation}\label{18.1}
   3\mathcal{H}^2 = a^2(\eta)\tilde{\rho}(\eta),
\end{equation}
and something similar for the acceleration equation \eqref{12}
\begin{equation}\label{18}
     a^2(\eta)p_r(\eta,r)=\mathcal{H}^2(\eta)-2\frac{\ddot{a}(\eta)}{a(\eta)}+\mathcal{P}(r)=a^2(\eta)\tilde{p}(\eta)+\mathcal{P}(r),
\end{equation}
which in turn results as 
\begin{equation}\label{18.2}
    \mathcal{H}^2-2\frac{\ddot{a}}{a}=a^2(\eta)\tilde{p}(\eta),
\end{equation}
where we have considered $\mathcal{P}(r)=-\frac{m_B(r)+M}{r^3}$. $\tilde{\rho}$, $\tilde{p}$ are the homogeneous quantities, thus, they depend only on $\eta$. Therefore $\mathcal{E}$ and $\mathcal{P}$ are deviations from homogeneous solutions due to inhomogeneities in energy density and pressure, respectively. Finally, plugging \eqref{18.1} and \eqref{18.2} into \eqref{13}, we get
\begin{equation}\label{18.3}
    \dot{\tilde{\rho}}+3\frac{\dot{a}}{a}(\tilde{\rho}+\tilde{p})=0,
\end{equation}
which is the FLRW continuity equation, with $\tilde{\rho}$ and $\tilde{p}$ representing the density and pressure. Notice that the scale factor is determined by $\tilde{\rho}$ and $\tilde{p}$ whereas the only effect of inhomogeneities is to produce the pressure $\mathcal{P}(r)$; meaning that the cosmological degrees of freedom are decoupled from inhomogeneities. Typically, in cosmology, we need an equation of state (EoS) that relates $p_r$ and $\rho$ to have a well-determined system. In the next section, we discuss the model of an EoS that is relevant for this work. 

\section{The Chaplygin model}
\label{sec:chaply}
For a description of an expanding universe, we need to find a component that transitions from dust-like to cosmological constant behavior. Some models are based on a scalar field to explain this transition, but such a description is not the only option. For an expanding universe, we need a positive energy density with negative pressure; the EoS of the Chaplygin gas model incorporates both characteristics. 

\subsection{The Chaplygin gas}
There are several variants of the Chaplygin model. We call the simplest EoS the Chaplygin gas, and it is of the form \cite{copeland2006dynamics, kamenshchik2001alternative}
\begin{equation}\label{19}
    p=-\frac{A}{\rho},
\end{equation}
if $\rho>0$, then $A$ is a constant that satisfies $A>0$. Substituting Eq. \eqref{19} into the usual continuity equation in the context of a FLRW spacetime, one gets 
\begin{equation}\label{20}
    \dot{\rho}+3\frac{\dot{a}}{a}\left(\frac{\rho^2-A}{\rho}\right)=0,
\end{equation}
a straightforward integration yields 
\begin{equation}\label{21}
    \rho(a)=\sqrt{A+\frac{B}{a^6}}
\end{equation}
with $B$ an integration constant. By analyzing this expression, for small values of $a$ and $\rho\sim\frac{\sqrt{B}}{a^3}$, which correspond to a dust-like behavior; for larger values of $a$, we make the following approximation: 
\begin{align*}
    \rho(a)=\sqrt{A}\sqrt{1+\frac{B}{Aa^6}},
\end{align*}
if $B/A\ll 1$ for $a\rightarrow 1$, then $\rho\sim\sqrt{A}$. Plugging this result into Eq. \eqref{19} we get $p=-\sqrt{A}$, allowing us to identify $\omega=-1$ under the assumption of a barotropic EoS between $\rho$ and $p$; this indicates a cosmological constant behavior with $\sqrt{A}$ in the role of the cosmological constant.

\subsection{The generalized Chaplygin gas EoS}
A simple generalization of the previous model is given by adding a new parameter, $\alpha$, in the following form \cite{bento2002generalized}
\begin{equation}\label{22}
    p=-\frac{A}{\rho^\alpha},
\end{equation}
this EoS is called the {\it generalized Chaplygin gas}; note that if $\alpha=1$, it reduces to the simple Chaplygin model. Doing the same process as described above, the continuity equation takes the form 
\begin{align*}
    \dot{\rho}+3\frac{\dot{a}}{a}\left(\frac{\rho^{\alpha+1}-A}{\rho^\alpha}\right)=0,
\end{align*}
integrating, we get the solution 
\begin{equation}\label{26}
    \rho(a)=\left[A+\frac{B}{a^{3(1+\alpha)}}\right]^{\frac{1}{1+\alpha}}.
\end{equation}
Before analyzing the limits $a\ll 1$ and $a\rightarrow 1$, we perform the following expansion for the obtained density
\begin{equation}\label{27}
    \rho(a)\approx A^{\frac{1}{1+\alpha}}\left[1+\frac{1}{1+\alpha}\frac{B}{A}a^{-3(1+\alpha)}\right],
\end{equation}
and for the pressure
\begin{equation}\label{28}
    p\approx -A^{\frac{1}{1+\alpha}}+\frac{\alpha}{1+\alpha}BA^{-\frac{\alpha}{1+\alpha}}a^{-3(1+\alpha)}. 
\end{equation}
With these two equations, we realize that in the intermediate regime between matter and a de Sitter universe, the behavior is characterized by a cosmological constant and a modified matter component represented by the barotropic EoS, $p=\alpha\rho$ \cite{CARDENAS2024101452}. 

As can be seen, the two different types of EoS discussed above are based on a simple enhancement in the way the cosmic fluid is described, providing a unified description of the dark sector. Recent studies have established the potential viability of black hole solutions in which this kind of fluid is involved. Although this description for the fluid is hypothetical, these findings not only suggest that such phenomena may plausibly exist in nature but also unveil new emergent physics related to thermodynamics phenomena of such configurations, see for instance Ref. \cite{Sekhmani_2024}; see also \cite{FATHI2024101598}. Since the availability of astrophysical observations is increasing day by day, this kind of scenario provides a theoretical laboratory to test the viability of possible candidates for the components of the dark sector.

\section{Discussion of the cosmological model}
\label{sec:disscu}
\subsection{Analytic solution for the Hubble parameter}
\begin{center}
\begin{figure}[htbp!]
    \includegraphics[width=0.5\textwidth]{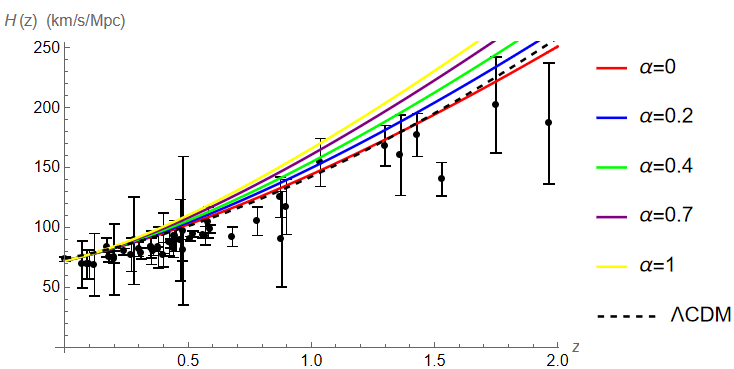}
    \includegraphics[width=0.5\textwidth]{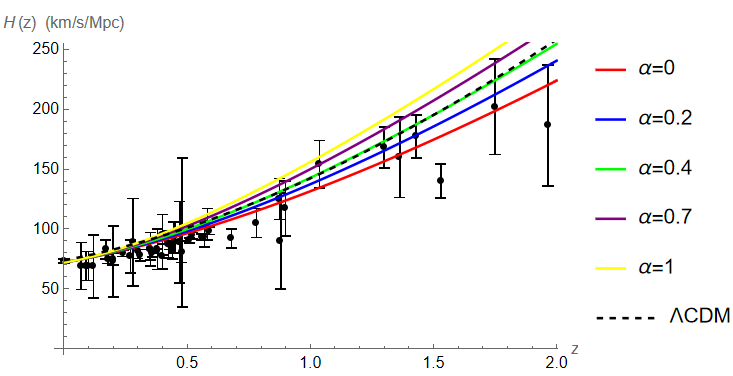}
	\caption{Comparison of $H(z)$ with $\Lambda$CDM considering observational Hubble data. Ratio $A/B=4/3$ for upper panel and $A/B=2$ for lower panel.}
	\label{fig:H1}
\end{figure}
\end{center}

By virtue of Eq. \eqref{18.3}, we can interpret $p$ and $\rho$ in Eq. \eqref{22} as $\tilde{p}$ and $\tilde{\rho}$. The solution of $\tilde{\rho}$ in terms of $a$ will be the one for the generalized Chaplygin in Eq. \eqref{26}. Inserting this expression in Eq. \eqref{18.1} we find 
\begin{equation}\label{H}
    3\mathcal{H}^2= a^2\left[A+\frac{B}{a^{3(1+\alpha)}}\right]^\frac{1}{1+\alpha},
\end{equation}
and from $H(t)=\frac{1}{a}\mathcal{H}(\eta)$, we obtain explicitly, as a function of the redshift, $z$, by means of the usual change of variable $1+z=a^{-1}$
\begin{equation}
    \label{hz1}
    \frac{H(z)}{H_0}\equiv E(z) =\left[\frac{A+B(1+z)^{3(1+\alpha)}}{A+B}\right]^{\frac{1}{2(1+\alpha)}},
\end{equation}
where $H_0$ is given by 
\begin{equation}\label{H0}
    H_0=\frac{1}{\sqrt{3}}\left(A+B\right)^\frac{1}{2(1+\alpha)}.
\end{equation}and $E(z)$ is the normalized Hubble parameter. In order to break the degeneracy between the parameters $A$ and $B$, we define $A/(A+B)=\Omega_{\text{de}}$, $B/(A+B)=\Omega_{\text{m}}$. Noticing that $\Omega_m+\Omega_{de}=1$, we can write $E(z)$ as follows 
\begin{equation}\label{Eom}
    E(z)=\left[\Omega_{\text{de}}+\Omega_{\text{m}}(1+z)^{3(1+\alpha)}\right]^\frac{1}{2(1+\alpha)}.
\end{equation}
In Figure \ref{fig:H1}, we show the behavior of $H(z)$ of the model against the Observational Hubble Data sample (OHD) taken from Refs. \cite{magana2018cardassian} and \cite{mukherjee2016acceleration} for different values of $\alpha$ and for two different values of the ratio of parameters $A$ and $B$. The $\Lambda$CDM curve is compared with the curves generated by our model for different parameter values. For the $4/3$ ratio, the curves resulting from the model lie above the $\Lambda$CDM curve except for the one with $\alpha=0$, whereas for the case $A/B= 2$, the $\Lambda$CDM model is almost reproduced with the value $\alpha =0.4$. 
We can also find an expression for $a(\eta)$ as an implicit function
\begin{eqnarray}\label{a}
    \eta &=& C_1+\frac{A^{-\frac{1}{2(1+\alpha)}}}{a}\\
    &\times & {}_2F_1\left[\frac{1}{2(1+\alpha)},\frac{1}{3(1+\alpha)},1+\frac{1}{3(1+\alpha)},-\frac{B}{Aa^{3(1+\alpha)}}\right], \nonumber
\end{eqnarray}
where $C_1$ is an integration constant and ${}_2F_1(a,b,c;x)$ is the Gauss hypergeometric function \footnote{A pedagogical discussion of the hypergeometric function can be found in \cite{seaborn2013hypergeometric}.}. Another important cosmological parameter is the deceleration parameter $q$, which is related to the Hubble parameter as follows:
\begin{equation}\label{5.11}
1+q = -\frac{\dot{H}}{H^2},
\end{equation}
where the dot indicates a derivative with respect to cosmological time \cite{DALARSSON2015237}. However, since $H$ is given as a function of the redshift, we can rewrite \eqref{5.11} as

\begin{equation}\label{5.12}
q = -1+(1+z) \frac{H^\prime}{H} ,
\end{equation}
where the prime denotes the derivative with respect to the redshift. In this case, we obtain 
\begin{equation}\label{dec}
    q(z)=-1+\frac{3}{2}\frac{B(1+z)^{3(1+\alpha)}}{\left[A+B(1+z)^{3(1+\alpha)}\right]}.
\end{equation}
\begin{center}
\begin{figure}[htbp!]
	\includegraphics[width=0.5\textwidth]{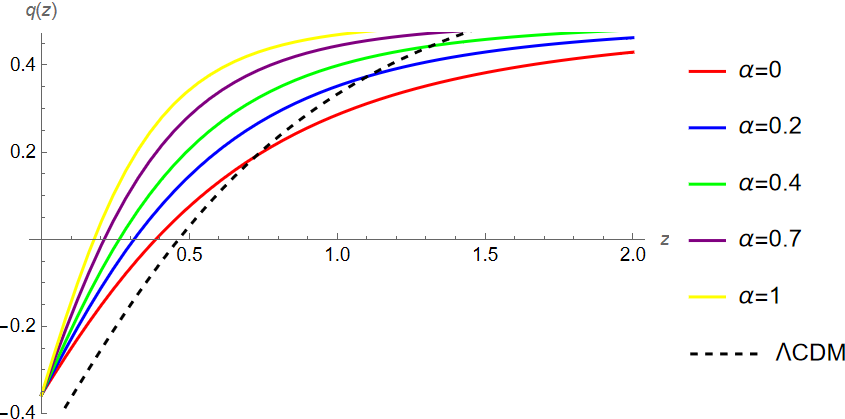}
    \includegraphics[width=0.5\textwidth]{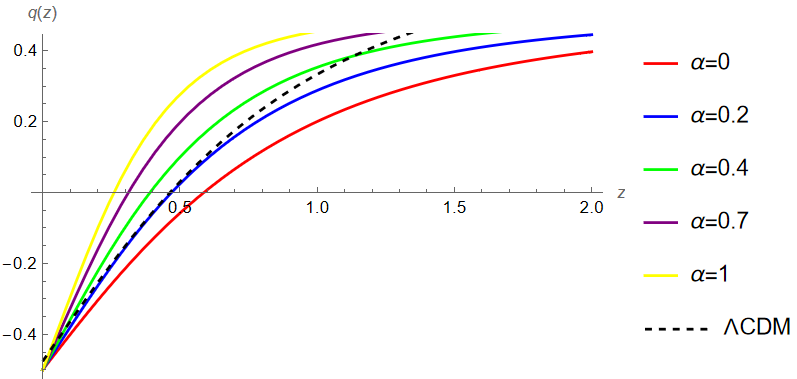}
	\caption{Comparison of $q(z)$ with $\Lambda$CDM. With ratio $A/B=4/3$ for the upper panel and $A/B=2$ for the lower one.}
	\label{fig:q1}
\end{figure}
\end{center}
In Figure \ref{fig:q1}, the model curves are compared to the $\Lambda$CDM curve for the same ratios as Figure \ref{fig:H1}. It can be seen that the transition from decelerated to accelerated expansion exists for every value of $\alpha$, so this model explains the accelerated expansion of the universe at late times. It can also be noted that the value of $\alpha$ is responsible for determining the moment in time -or the value of $z$- at which the curve changes sign. Consistent with Fig. \ref{fig:H1}, the model resembles the standard case more accurately for the ratio $A/B=2$.  

\subsection{Approximation for late universe}
The late universe is a very important cosmological limit; therefore, we focus on the task of finding an expression that is valid in that time regime. We take as a starting point equation \eqref{18.3}, where we identify $\tilde{\rho}$ and $\tilde{p}$ as the two quantities that will be related by means of the EoS described by Eq. \eqref{22}. To solve the differential equation \eqref{18.1}, we can perform the first-order expansion given by \eqref{27}, but for this expansion to be valid, the condition $B/A\ll 1$ must be satisfied. The differential equation can be approximated as:
\begin{equation}\label{5.6}
    \dot{a}\simeq \frac{a^2}{\sqrt{3}}A^{\frac{1}{2(1+\alpha)}}\left[1+\frac{1}{2(1+\alpha)}\frac{B}{Aa^{3(\alpha+1)}}\right],
\end{equation}
whose solution is given in terms of the following implicit function with respect to conformal time as:
\begin{equation}\label{5.7}
\eta+C_1=\frac{a^{2+3\alpha}\beta}{\kappa(2+3\alpha)}{}_2F_1\left[1,\frac{2+3\alpha}{3(1+\alpha)},\frac{5+6\alpha}{3(1+\alpha)},-\frac{a^{3(1+\alpha)}}{\kappa}\right]
\end{equation}
where $\kappa$ and $\beta$ are constants defined as $\kappa=\frac{1}{2(1+\alpha)}\frac{B}{A}$ and $\beta=3A^{-\frac{1}{1+\alpha}}$ and $C_1$ is an integration constant. If we differentiate this latter equation with respect to conformal time, we obtain:
\begin{eqnarray}
1 &=& \beta\frac{a^{1+3\alpha}}{\kappa}\dot{a}\Biggl\{{}_2F_1\left[1,\frac{2+3\alpha}{3(1+\alpha)},\frac{5+6\alpha}{3(1+\alpha)},-\frac{a^{3(1+\alpha)}}{\kappa}\right] \label{5.8} \\
 	&-&\frac{3(1+\alpha)}{\kappa(5+6\alpha)}a^{3(\alpha+1)}{}_2F_1\left[2,\frac{5+6\alpha}{3(1+\alpha)},\frac{8+9\alpha}{3(1+\alpha)},-\frac{a^{3(1+\alpha)}}{\kappa}\right]\Biggr\}, \nonumber 
\end{eqnarray}    
Therefore, with the use of the expression \eqref{5.8}, it is easy to find the conformal Hubble parameter, $\mathcal{H}=\frac{\dot{a}}{a}$, as a function of the scale factor, as follows:
\begin{widetext}
 \begin{equation}\label{5.9}
 	\mathcal{H}(a)=\frac{\kappa}{\beta a^{2+3\alpha}\Biggl\{{}_2F_1\left[1,\frac{2+3\alpha}{3(1+\alpha)},\frac{5+6\alpha}{3(1+\alpha)},-\frac{a^{3(1+\alpha)}}{\kappa}\right]-\frac{3(1+\alpha)}{\kappa(5+6\alpha)}a^{3(\alpha+1)}{}_2F_1\left[2,\frac{5+6\alpha}{3(1+\alpha)},\frac{8+9\alpha}{3(1+\alpha)},-\frac{a^{3(1+\alpha)}}{\kappa}\right]\Biggr\}}.
 \end{equation}
 \end{widetext}
 It is convenient to express the Hubble parameter in terms of the redshift $z$, since this quantity is our physical observable. By making this variable change, equation \eqref{5.9} can be rewritten as:
 \begin{widetext}
  \begin{equation}\label{5.10}
 	H(z)=\frac{\kappa(1+z)^{2+3\alpha}}{\beta\Biggl\{{}_2F_1\left[1,\xi,\xi+1,-\frac{1}{\kappa(1+z)^{3(1+\alpha)}}\right]-\frac{3(1+\alpha)}{\kappa(5+6\alpha)(1+z)^{3(\alpha+1)}}{}_2F_1\left[2,\xi+1,\xi+2,-\frac{1}{\kappa(1+z)^{3(1+\alpha)}}\right]\Biggr\}},
 \end{equation}
 \end{widetext}
 where, for simplicity in the notation, we have defined the constant $\xi:=\frac{2+3\alpha}{3(1+\alpha)}$.
\begin{center}
\begin{figure}[htbp!]
 	\includegraphics[width=0.5\textwidth]{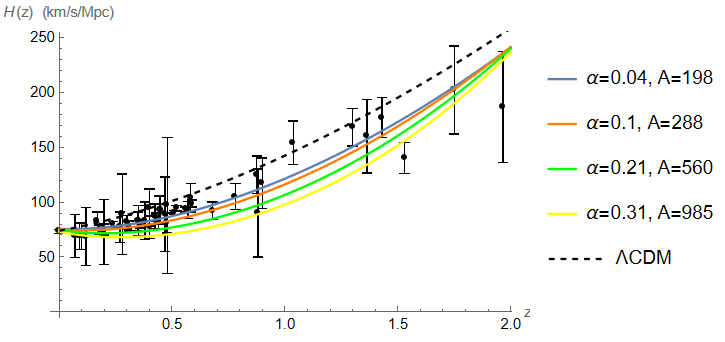}
    \includegraphics[width=0.5\textwidth]{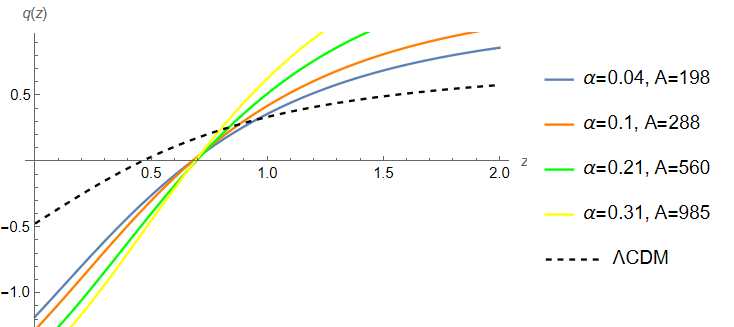}
 	\caption{$H(z)$ and $q(z)$ compared to the $\Lambda$CDM for different values of parameters $\alpha$ and $A$.}
 	\label{fig:Ha}
\end{figure}
\end{center}
In the upper panel of Figure \ref{fig:Ha}, we plot the parameter $H(z)$ of the model against observational data for different values of the parameters $\alpha$ and $A$. It may seem exaggerated for the parameter $A$ to take such large values; however, we must remember the approximation that was made in order to calculate analytical solutions and the condition that this parameter must satisfy, and we compare to the $\Lambda$CDM model. Additionally, in the lower panel of Figure \ref{fig:Ha} we show the deceleration parameter obtained from (\ref{5.10}). The model curves are compared to the $\Lambda$CDM curve. It can be observed that, in this case, our model curves are much more pronounced than the $\Lambda$CDM case, from which we can infer a more abrupt transition from deceleration to acceleration, as well as a much greater acceleration of the universe than that described by the standard model. Similarly, we note that the model curves intersect the $z$-axis earlier than the $\Lambda$CDM model, which indicates that the transition from deceleration to acceleration begins earlier than predicted by the $\Lambda$CDM.

\subsubsection{Connection with the analytic solution for $H$}
Now, recalling the differential equation \eqref{5.6} and from the definition of $H(t)$ we get
\begin{equation}\label{Htlate}
    H(a)=\frac{A^\frac{1}{2(1+\alpha)}}{\sqrt{3}}\left[1+\frac{B}{2A(1+\alpha)a^{3(1+\alpha)}}\right]
\end{equation}
corresponding to a limiting case of Eq. \eqref{5.10}, we get the following solution to Eq. \eqref{5.6} 
\begin{equation}\label{late}
    \frac{H(z)}{H_0}=E(z)=\left[\frac{2A(1+\alpha)+B(1+z)^{3(1+\alpha)}}{2A(1+\alpha)+B}\right],
\end{equation}
where $H_0$ is given by 
\begin{equation}\label{H0late}
    H_0=\frac{A^\frac{1}{2(1+\alpha)}}{\sqrt{3}}\left[1+\frac{B}{2A(1+\alpha)}\right],
\end{equation}
as can be seen, these results can be obtained from Eqs. (\ref{hz1}) and (\ref{H0}) under the consideration of $B/A \ll 1$. 

\section{Observations and the viability of the model}
\label{sec:obs}
\subsection{Distance modulus and redshift relation}
In this section, we perform some preliminary tests of the model, which may serve as a foundation for more comprehensive analyses in the future, incorporating more recent observational datasets and potentially employing more sophisticated methods.
Before performing a statistical analysis, we compare the theoretical predictions of the model with observational data by studying the distance modulus--redshift relation. We firstly computed the luminosity distance as a function of redshift, given by:
\begin{equation}
d_L(z) = a_0(1+z)S_k(\chi(z)),
\end{equation}
where $a_0$ is the present scale factor and $S_k$ is the curvature-dependent comoving angular diameter distance. Since we assume a spatially flat universe, the comoving distance simplifies to:
\begin{equation}
S_k(D_M(z)) = D_M(z) = \int_0^z \frac{c\,dz}{H(z)} = \frac{c}{H_0} \int_0^z \frac{dz}{E(z; \Omega_m, \alpha)},
\end{equation}
where $c$ is the speed of light and the function $E(z; \Omega_m, \alpha)$ is defined by the expression given in Eq.~(\ref{hz1}). The notation $E(z; \Omega_m, \alpha)$ implies that the normalized Hubble parameter is a function of the redshift $z$ and contains the free parameters $\Omega_m$ and $\alpha$. The numerical evaluation of the integral was carried out using the trapezoidal method implemented in our own Python code. Given that the truncation error of this method is of second order in the step size, the numerical accuracy obtained is sufficiently reliable for our analysis. Once the luminosity distance was determined, we computed the distance modulus using the standard relation:
\begin{equation}
\mu(z) \equiv m - M = 5\log_{10}\left(\frac{d_L(z)}{1\,\mathrm{pc}}\right) - 5.
\end{equation}

\begin{figure}[htbp!]
	\centering
	\includegraphics[width=0.5\textwidth]{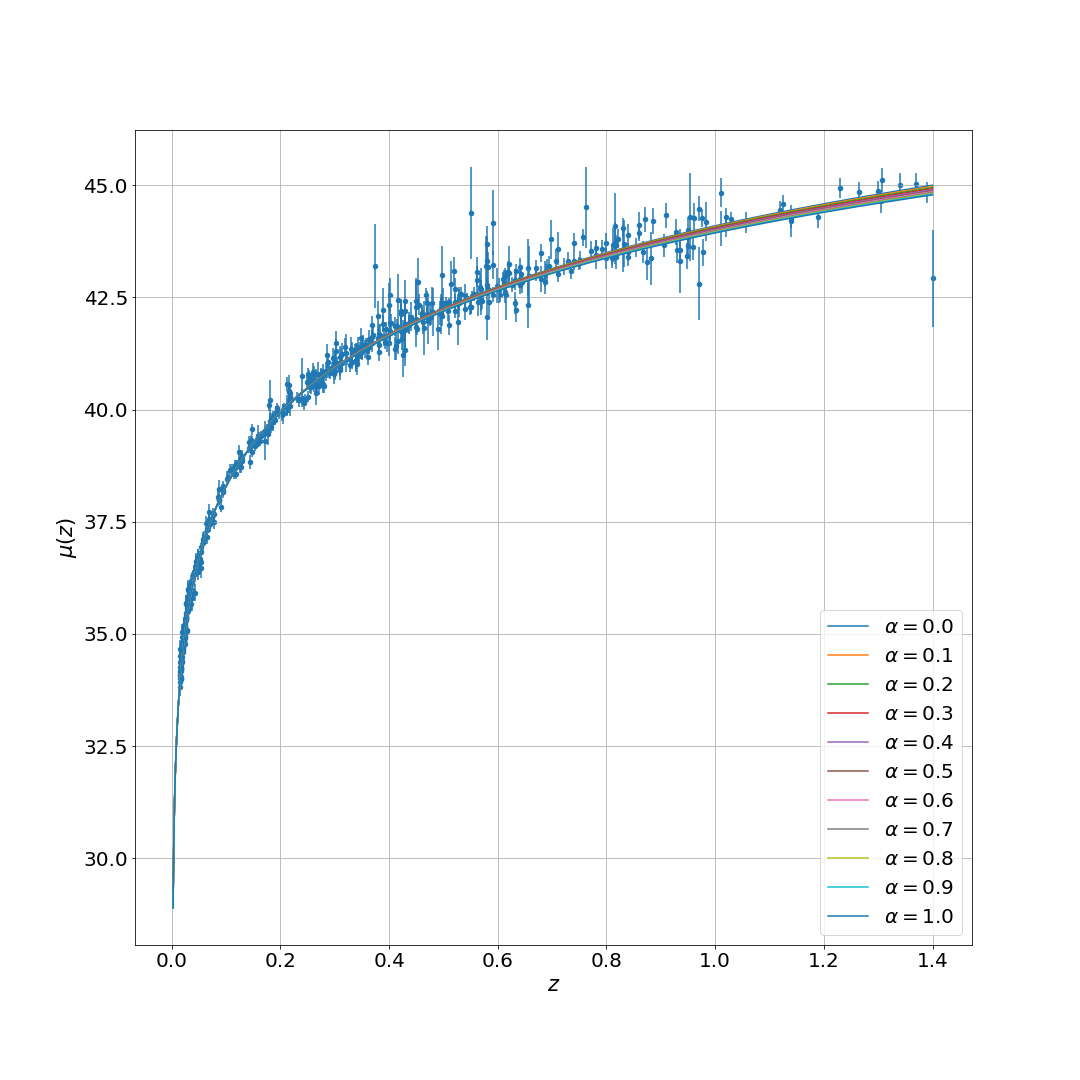}
	\caption{Predicted distance modulus as function of red-shift from our model overlapped to the observed modulus for Union 2.2 supernovae. The values of the parameters for this model corresponds to $\Omega_m=0.3$, $H_0 =70\,\mathrm{km/s/Mpc}$ and $0\leq\alpha\leq1$.}
	\label{fig:mu}
\end{figure}

Figure~\ref{fig:mu} shows the behavior of $\mu(z)$ as a function of redshift for different values of the parameter $\alpha$, which characterizes the deviation from standard behavior in the generalized Chaplygin gas equation of state. The observational data points, along with their associated uncertainties, correspond to the Union 2.2 supernovae compilation \cite{Amanullah_2010}.

It is evident that the theoretical curves corresponding to different values of $\alpha$ exhibit a very similar evolution throughout the redshift interval considered, especially at low and intermediate redshifts. As $\alpha$ increases, the model shows a slight deviation from the standard scenario; however, all curves remain within the uncertainty range of the observational data. This suggests a degeneracy in the effect of $\alpha$ on the distance modulus, implying that different values of this parameter can produce cosmological histories that are indistinguishable using supernova data alone.

\subsection{Test using BAO}

Before proceeding with the Bayesian analysis, we extend our comparison of theoretical predictions with observational data by studying the baryon acoustic oscillation (BAO) distance observables, specifically the transverse comoving distance \(D_M(z)\) and the volume-averaged distance \(D_V(z)\). These observables are crucial for constraining the expansion history of the Universe beyond the distance modulus relation.

As shown previously, the generalized dark energy model introduces a parameter \(\alpha\) that modifies the Hubble expansion rate as
\begin{equation}
E(z; \Omega_m, \alpha) = \left[ \Omega_{\mathrm{de}} + \Omega_m (1+z)^{3(1+\alpha)} \right]^{\frac{1}{2(1+\alpha)}},
\end{equation}
where \(\Omega_{\mathrm{de}} = 1 - \Omega_m\), assuming spatial flatness. 

The standard \(\Lambda\mathrm{CDM}\) model corresponds to the particular case \(\alpha = 0\). The transverse comoving distance is computed by
\begin{equation}
D_M(z) = \frac{c}{H_0} \int_0^z \frac{dz'}{E(z'; \Omega_m, \alpha)},
\end{equation}
while the sound horizon at the baryon drag epoch \(z_d\) is
\begin{equation}
r_s(\alpha) = \int_{z_d}^\infty \frac{c_s(z)}{H_0 E(z; \Omega_m, \alpha)}\, dz,
\end{equation}
with the sound speed \(c_s(z)\) in the photon-baryon fluid, given by
\begin{equation}
c_s(z) = \frac{c}{\sqrt{3(1 + R(z))}}, \quad R(z) = \frac{3 \Omega_b}{4 \Omega_\gamma (1+z)}.
\end{equation}

\begin{figure}[htbp!]
    \centering
    \includegraphics[width=0.5\textwidth]{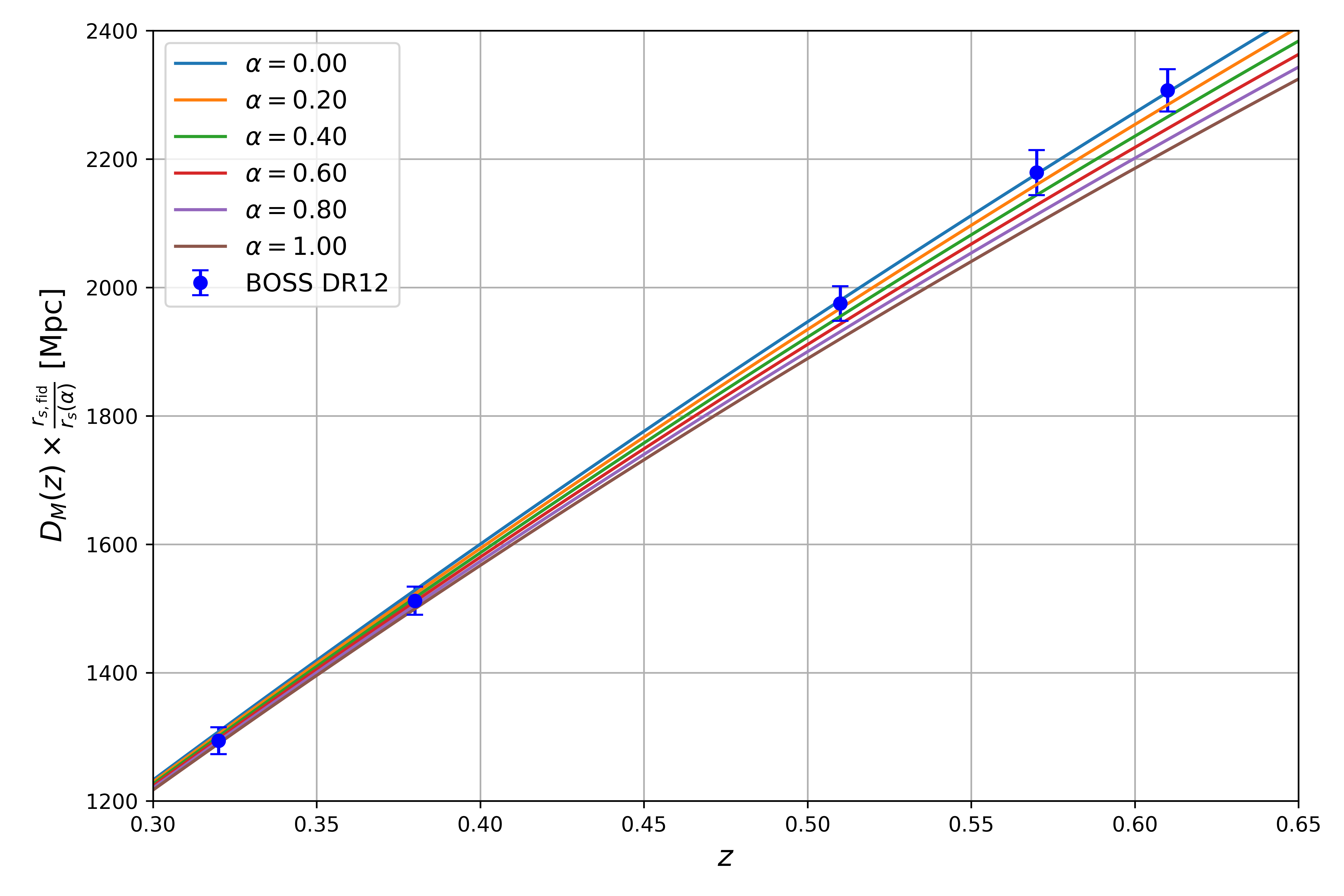}
    \caption{Traverse comoving distance predicted in a sample of models with various values of $\alpha$ along with BAO distance measurements inferred from the 12th release of SDSS reported in \cite{BOSS:2016wmc}}
    \label{fig:DM}
\end{figure}

The volume-averaged distance \(D_V(z)\), which combines radial and transverse BAO scales, is defined as
\begin{equation}
D_V(z) = \left[ (1+z)^2 D_M^2(z) \frac{c z}{H_0 E(z; \Omega_m, \alpha)} \right]^{1/3}.
\end{equation}

\begin{figure}[htbp!]
    \centering
    \includegraphics[width=0.5\textwidth]{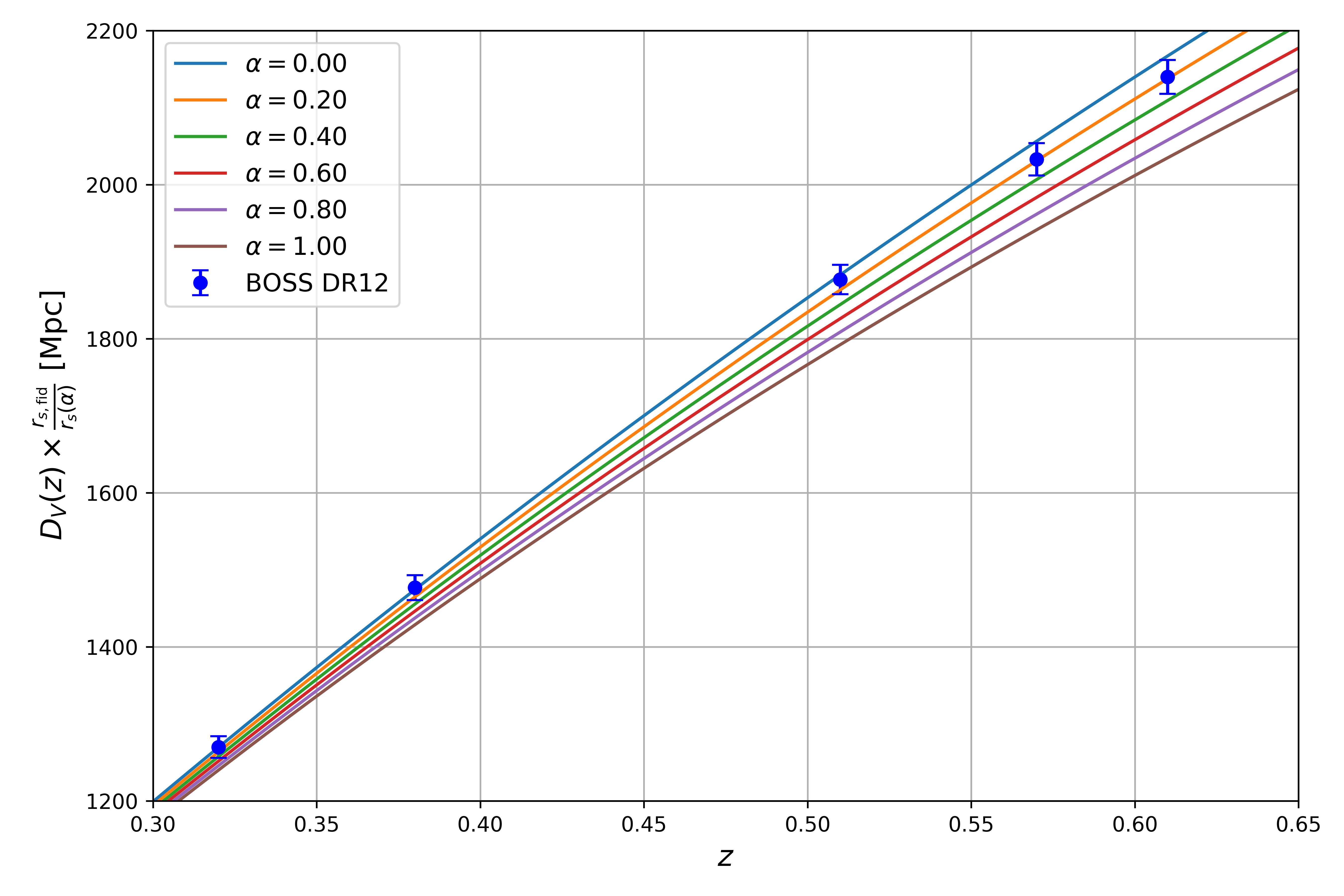}
    \caption{Volume-averaged distance predicted in a sample of models with various values of $\alpha$ along with BAO distance measurements inferred from the 12th release of SDSS reported in \cite{BOSS:2016wmc}}
    \label{fig:DV}
\end{figure}

In our computations, cosmological parameters were set close to Planck 2018 values \cite{2020}: \(H_0 = 70\, \mathrm{km/s/Mpc}\), \(\Omega_b = 0.0486\), \(\Omega_\gamma = 5 \times 10^{-5}\), and the fiducial sound horizon \(r_{s,\mathrm{fid}} = 147.78\, \mathrm{Mpc}\). The drag redshift was fixed at \(z_d = 1059\).

The theoretical quantities compared with the BAO data from SDSS DR12 (BOSS) \cite{BOSS:2016wmc} are
\begin{equation}
D_M(z) \times \frac{r_{s,\mathrm{fid}}}{r_s(\alpha)} \quad \text{and} \quad D_V(z) \times \frac{r_{s,\mathrm{fid}}}{r_s(\alpha)},
\end{equation}
evaluated at those observational redshifts, which can be seen in Figure~\ref{fig:DM} and Figure~\ref{fig:DV}, respectively. By varying the parameter \(\alpha\) from 0 (the \(\Lambda\mathrm{CDM}\) case) up to 1, we analyzed how the model predictions for these distances shift relative to the data points.

This comparison provides a complementary probe to the distance modulus test, revealing the sensitivity of BAO observables to the dark energy parameter \(\alpha\). It also allows for tighter constraints on the model parameters by including multiple cosmological distance measures in the subsequent Bayesian analysis.

Although this qualitative comparison reveals that the model provides a good fit to the data even for relatively large values of $\alpha$, a detailed statistical approach is required to determine the most probable values of the model parameters. For this purpose, we now proceed with a Bayesian analysis based on the same observational datasets.

\subsection{Bayesian analysis}

The cosmological model considered in this work, characterized by an anisotropic background dominated by a generalized polytropic Chaplygin-type fluid, was constrained using observational data from the Union 2.2 \cite{Amanullah_2010} supernovae compilation and SDSS DR12 from BOSS \cite{BOSS:2016wmc} through a Bayesian statistical analysis with 25\% acceptance probability. The model includes three free parameters: the matter density \( \Omega_m \), the Hubble constant \( H_0 \), and a phenomenological parameter \( \alpha \) that controls deviations from the standard expansion history. The best-fit values of the model parameters, along with their associated $1\sigma$ uncertainties, are presented in Table~\ref{tab:table1}. Additionally, the  figures~\ref{fig:BAO},~\ref{fig:UNION} and~\ref{fig:BAOUNION} given in the appendix (\ref{sec:separated}) display the marginalized posterior distributions and the confidence contours (68\% and 95\%) for each pair of parameters using SDSS DR 12, Union 2.2 and SDSS DR 12 $+$ Union 2.2. Figure~\ref{fig:PD} shows a comparison using the three posterior distributions.

The BAO likelihood was constructed using the transverse comoving distance \( D_M(z) \) and the volume-averaged distance \( D_V(z) \), both normalized by a fiducial sound horizon scale. For the supernovae, the theoretical distance modulus \( \mu(z) \) was computed from the luminosity distance \( d_L(z) = (1+z)D_M(z) \), under the assumption of spatial flatness. Numerical integrations required for the computation of each observable were performed using adaptive quadrature methods implemented in Python.

We employed the Metropolis-Hastings algorithm to explore the parameter space, running multiple independent Markov chains with distinct initial conditions to ensure good mixing and convergence. The posterior distributions were analyzed using the \texttt{getdist} package, which provided the marginalized intervals and triangular plots showing the parameter degeneracies. Finally, we compared the constraints obtained from each dataset individually with those from the combined analysis, highlighting the improvement in parameter estimation—particularly for the parameter \( \alpha \)—when both datasets are used simultaneously.

\subsection{Discussion}
The results indicate that the matter density parameter, $\Omega_m$, lies slightly above the value favored by the $\Lambda$CDM model as constrained by Planck \cite{2020}, though it remains consistent within $1\sigma$. The Hubble constant is constrained in excellent agreement with the value inferred by Planck, suggesting that the present model does not exacerbate the well-known Hubble tension and may be compatible with early-universe observations. The parameter $\alpha$, which characterizes deviations from perfect fluid behavior and controls the exotic dynamics of the generalized Chaplygin fluid, points to a non-trivial contribution to cosmic acceleration that differs from a pure cosmological constant.

\begin{center}
\begin{figure}[htbp!]
  \includegraphics[width=0.6\textwidth]{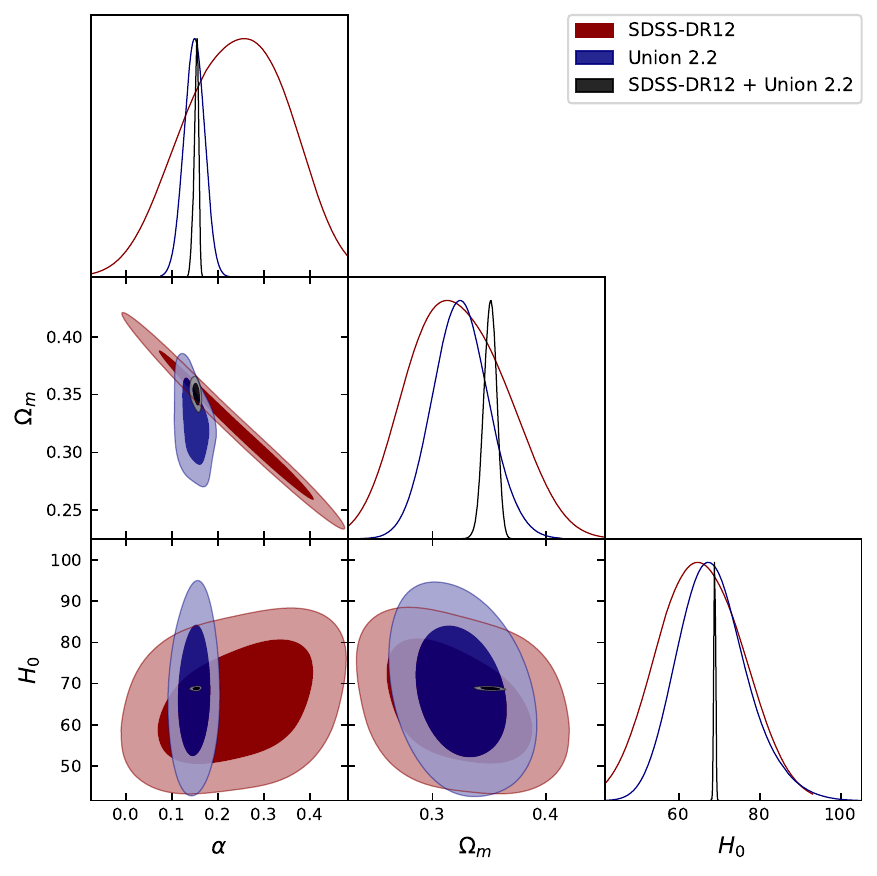}
\caption{Comparison of the 1D and 2D marginalized posterior distributions for the parameters $\Omega_m$, $H_0$ and $\alpha$ for SDSS DR 12, UNION 2.2 and SDSS DR12 + UNION 2.2}
\label{fig:PD}
\end{figure}
\end{center}

The triangular plots provide insight into the structure of the parameter space, revealing a negative correlation between $\Omega_m$ and $H_0$, a feature commonly found in cosmological models due to their competing effects on the expansion rate. A moderate correlation is also observed between $H_0$ and $\alpha$, indicating a degree of degeneracy between the expansion dynamics and the equation of state of the dark sector. The posterior distributions are well-behaved, showing no signs of multimodality or extended non-Gaussian tails, which supports the statistical robustness of the parameter estimates. As commented above, our main results are explicitly shown in Fig. (\ref{fig:PD}) and summarized in Table (\ref{tab:table1}).

Overall, these results suggest that the proposed model is a viable alternative to the standard $\Lambda$CDM framework, capable of accommodating observational data without the explicit inclusion of a cosmological constant. A quantitative comparison with $\Lambda$CDM through Bayesian evidence or information criteria would further clarify whether the introduction of the parameter $\alpha$ is statistically justified by a significant improvement in the data fit. In the appendix (\ref{sec:separated}) the results for the 1D and 2D posterior distributions are presented separately in order to visualize more clearly the effects of the consideration of different datasets on the confidence regions obtained for the parameters of the model. 
\onecolumngrid
\twocolumngrid

\onecolumngrid
\begin{center}
\begin{table}[htbp!]
\centering
\begin{tabular}{lccc}
\toprule
\textbf{Parameter} & \textbf{Union 2.2} & \textbf{SDSS-DR12} & \textbf{Combined} \\
\midrule
$\alpha$     & $0.15013^{+0.01669}_{-0.02186}$ & $0.2398^{+0.1203}_{-0.1085}$ & $0.1529^{+0.0058}_{-0.0038}$ \\
$\Omega_m$   & $0.32506^{+0.02128}_{-0.02411}$ & $0.3224^{+0.0402}_{-0.0465}$ & $0.3508^{+0.0058}_{-0.0054}$ \\
$H_0$        & $67.76971^{+8.35813}_{-8.02357}$  & $65.2119^{+10.3672}_{-10.6208}$  & $68.8179^{+0.2447}_{-0.2209}$ \\
\bottomrule
\end{tabular}
\caption{Adjusted parameters with 1$\sigma$ confidence level.}
\label{tab:table1}
\end{table}
\end{center}
\twocolumngrid

\subsection{Quantitative analysis}

Table~\ref{tab:lcdm-vs-alpha-extended} presents the statistical comparison between the concordance $\Lambda$CDM model ($k=2$) and the $\alpha$-extended model ($k=3$). In this context, $N$ denotes the number of data points, $k$ the number of free parameters, $\chi^2$ the chi-squared statistic, and $\chi^2_\nu=\chi^2/(N-k)$ the reduced chi-squared, which evaluates the fit quality per degree of freedom. By construction, $\chi^2$ is expected to be of order $\nu=N-k$, so that $\chi^2_\nu\approx 1$ is the reference for a statistically consistent fit. Values of $\chi^2_\nu \ll 1$ suggest that the model fits comfortably within the quoted uncertainties, possibly reflecting conservative error bars, while $\chi^2_\nu \gg 1$ would indicate tension between model and data.
In addition, we employ two information-theoretic criteria to compare models of different complexity. The Akaike Information Criterion (AIC) is defined as $\mathrm{AIC}=\chi^2+2k$, while the Bayesian Information Criterion (BIC) is defined as $\mathrm{BIC}=\chi^2+k\ln N$. Both include the chi-squared term to reward goodness of fit, but they also introduce a cost that increases with the number of free parameters $k$, thereby discouraging overfitting. In the case of AIC the cost grows linearly with $k$, whereas in BIC it grows as $k\ln N$. This means that for small datasets the two criteria behave similarly, but for large datasets BIC penalizes additional parameters more strongly than AIC. For convenience, we report not only the individual AIC and BIC values but also the differences $\Delta\mathrm{AIC}$ and $\Delta\mathrm{BIC}$ (extended minus concordance). When two different are compared, the favored one by observations has smaller values for $\chi^{2}$ and BIC. In general if $\Delta$BIC between models lies in the interval $2-6$ indicates evidence against the model with higher BIC, a difference given between $6-10$ reveals strong evidence against the model and a difference higher than $10$ must be taken as very strong evidence against the model with higher BIC \cite{c4048c8f-6ca9-3965-96a3-653ab8996955}. 
For the BAO dataset ($N=5$), both models yield nearly identical results: $\chi^2_{\Lambda\mathrm{CDM}}=2.11$ and $\chi^2_{\alpha}=2.09$, with $\chi^2_\nu=0.70$ in both cases (given $\nu=3$ and $\nu=2$). Since the expected reference is $\chi^2_\nu\approx 1$, these values indicate that the BAO data are reproduced reasonably well within the observational uncertainties. The differences $\Delta\mathrm{AIC}=+1.98$ and $\Delta\mathrm{BIC}=+2.19$ arise mainly from the cost associated with the extra parameter, leaving the two descriptions essentially comparable in absolute terms.

For the Union~2.2 supernovae dataset ($N=80$), the outcome is similar: $\chi^2_{\Lambda\mathrm{CDM}}=40.64$ and $\chi^2_{\alpha}=40.67$, both with $\chi^2_\nu=0.52$ (for $\nu=78$ and $77$). Compared to the expected $\chi^2_\nu\approx 1$, these values suggest that the models fit the data adequately, though the relatively low values may also reflect conservative error estimates. The small positive shifts $\Delta\mathrm{AIC}=+2.03$ and $\Delta\mathrm{BIC}=+3.58$ represent mild costs for the extended model, which otherwise performs comparably to the concordance case.

For the combined dataset ($N=85$), the same picture emerges: $\chi^2_{\Lambda\mathrm{CDM}}=42.75$ and $\chi^2_{\alpha}=42.76$, with $\chi^2_\nu=0.52$ in both cases ($\nu=83$ and $82$). Once again, the fit quality is acceptable and the information-criterion differences $\Delta\mathrm{AIC}=+2.01$ and $\Delta\mathrm{BIC}=+4.45$ point to a modest preference for the concordance model.

Overall, these results show that the $\Lambda$CDM model is slightly statistically preferred by the current datasets, as the small improvement in $\chi^2$ does not compensate for the additional parameter under the $\mathrm{AIC}/\mathrm{BIC}$ costs. Nevertheless, the fact that its absolute fit quality, measured through $\chi^2$ and $\chi^2_\nu$, remains comparable to that of the concordance model means that such extensions are competitive and cannot be ruled out. While $\Lambda$CDM continues to be slightly favored for its simplicity, exploring extended frameworks remains valuable, since future datasets with improved precision and redshift coverage may reveal subtle deviations that current observations are not yet able to detect. Therefore, we can consider the $\alpha$-extended model as a good theoretical framework to describe the recent accelerated expansion of the universe without the inclusion of a cosmological constant or some type of dark energy.
\onecolumngrid
\begin{center}
\begin{table}[htbp!]
\centering
\scriptsize
\resizebox{\textwidth}{!}{%
\begin{tabular}{l
S[table-format=3.0]
S[table-format=1.0]
S[table-format=5.2]
S[table-format=1.2]
S[table-format=5.2]
S[table-format=5.2]
S[table-format=1.0]
S[table-format=5.2]
S[table-format=1.2]
S[table-format=5.2]
S[table-format=5.2]
S[table-format=+4.2]
S[table-format=+4.2]
}
\toprule
Dataset & {$N$} & {$k_{\Lambda\mathrm{CDM}}$} & {$\chi^2_{\Lambda\mathrm{CDM}}$} & {$\chi^2_{\nu,\Lambda\mathrm{CDM}}$} & {AIC$_{\Lambda\mathrm{CDM}}$} & {BIC$_{\Lambda\mathrm{CDM}}$} & {$k_{\alpha}$} & {$\chi^2_{\alpha}$} & {$\chi^2_{\nu,\alpha}$} & {AIC$_{\alpha}$} & {BIC$_{\alpha}$} & {$\Delta$AIC} & {$\Delta$BIC} \\
\midrule
SDSS-DR12 (BAO)    & 5  & 2 & 2.11  & 0.70 & 6.11  & 6.51  & 3 & 2.09  & 0.70 & 8.09  & 8.70  & +1.98 & +2.19 \\
Union 2.2 (SNe)    & 80 & 2 & 40.64 & 0.52 & 44.64 & 49.21 & 3 & 40.67 & 0.52 & 46.67 & 52.79 & +2.03 & +3.58 \\
Combined (BAO+SNe) & 85 & 2 & 42.75 & 0.52 & 46.75 & 51.64 & 3 & 42.76 & 0.52 & 48.76 & 56.08 & +2.01 & +4.45 \\
\bottomrule
\end{tabular}
}
\caption{Statistical comparison between $\Lambda$CDM ($k=2$) and the $\alpha$-extended model ($k=3$) using BAO, SNe, and their combination. The values for $\chi^2$, reduced chi-squared $\chi^2_\nu$, AIC, BIC, and the differences $\Delta$AIC/$\Delta$BIC are presented.}
\label{tab:lcdm-vs-alpha-extended}
\end{table}
\end{center}
\twocolumngrid

\begin{center}
\begin{figure}[htbp!]
    \centering
    \includegraphics[width=0.65\textwidth]{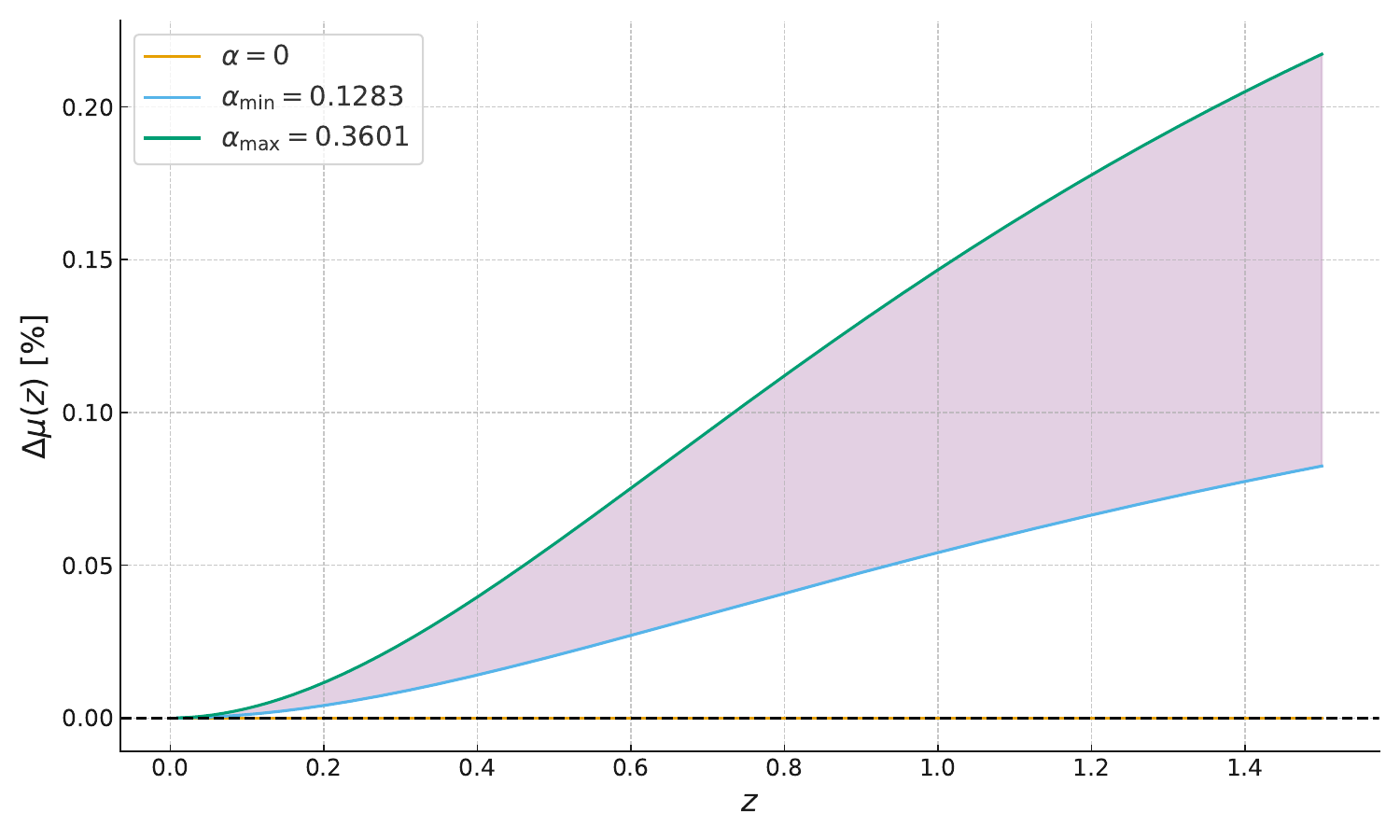}
    \caption{Percentage difference $\Delta \mu(z)[\%]$ as function of the redshift for upper and lower best fit adjusted values of the $\alpha$ parameter.}
    \label{fig:percentage}
\end{figure}
\end{center}

To end this section, the deviations between the extended scenario and the concordance model can be quantified by computing the relative percentage difference of the distance modulus, defined as
\begin{equation}
\Delta \mu(z)[\%] \;=\; 
\frac{|\mu_{\alpha}(z)-\mu_{\Lambda CDM}(z)|}{\mu_{\Lambda CDM}(z)} \times 100 ,
\end{equation}
where $\mu_{\alpha}(z)$ corresponds to the extended model with the extra parameter $\alpha$, and $\mu_{\Lambda CDM}(z)$ denotes the prediction of the standard $\Lambda$CDM cosmology. 
This quantity expresses, in percentage terms, how much the extended model departs from the concordance case across the redshift range. 
In Fig. (\ref{fig:percentage}) the shaded region shows that $\Delta \mu(z)$ remains at the percent level under the consideration of the best fit values obtained for the parameter $\alpha$, indicating that the extended framework reproduces the overall behavior of the concordance model and at the recent past and at present time of cosmic evolution are practically indistinguishable, while still allowing for small systematic departures that may become significant in precision cosmology analyses. The introduction of the extra parameter $\alpha$ leads to percent-level deviations in the distance modulus with respect to the concordance model, suggesting that the extended framework is broadly consistent with the standard scenario while leaving room for potentially relevant differences in high-precision studies.

\onecolumngrid
\twocolumngrid

\section{Conclusions}
\label{sec:concl}
Standard cosmological models, while successful, rely on the simplifying assumption of perfect homogeneity. To move towards a more physically realistic description of our universe, we have developed our analysis within the context of an inhomogeneous and anisotropic spacetime model. The starting point of this analysis was the metric given by \eqref{1}. The inhomogeneity is encoded in the metric functions $f(r)$ and $\gamma(r)$. We chose these functions to be radially dependent to ensure that their contributions naturally vanish at the appropriate cosmological limit, ensuring consistency with the FLRW metric on large scales. The anisotropy arises naturally from distinct pressure components in the energy-momentum tensor \eqref{2}. As discussed previously, the cosmological degrees of freedom decouple from inhomogeneity; therefore, the behavior of such quantities is described by the standard dynamical equations. Additionally, the fluid's profile is described by modeling its behavior throughout the generalized Chaplygin EoS, which represents a more robust description than the barotropic case, since it can describe the dark sector in a unified way under certain considerations. A key feature of the model is its dependence on a single free parameter, $\alpha$, which is introduced through the EoS.

As a first critical test, the model proves its viability under the following criteria: the resulting Hubble parameter agrees with observational data; crucially, the deceleration parameter transitions from positive to negative, correctly reproducing the observed shift from a decelerating to an accelerating phase of cosmic expansion. As shown in our statistical analysis, by incorporating inhomogeneity from first principles, this model offers a robust framework for testing against further observational data, such as luminosity distance, and holds the potential to shed new light on fundamental processes, such as the growth of large-scale structures. We leave this subject open for future investigation.

\begin{acknowledgments}
G.A.P. was supported by SECIHTI through the program {\it Estancias Posdoctorales por México 2023(1)}. M.C. work was partially supported by S.N.I.I. (SECIHTI-M\'exico). M.F. is supported by Universidad Central de Chile through project No. PDUCEN20240008.  
J.R.V. is partially supported by Centro de F\'isica Teórica de Valparaíso (CeFiTeV). G.A.P. thanks the warm hospitality of the colleagues at the Instituto de Física y Astronomía of the Universidad de Valparaíso, Chile, where part of this work was carried out.
\end{acknowledgments}

\appendix
\section{The Kretschmann scalar}
As a case of interest, we show that from the spacetime metric (\ref{1}), the Kretschmann scalar can be written as
\begin{widetext}
\begin{eqnarray}
\mathcal{K}=R_{\mu\nu\alpha\beta}R^{\mu\nu\alpha\beta} &=& \frac{e^{-2 \gamma}}{4 r^4 a^8 f^2}\Biggl\{
-8 e^{\gamma}f r^3 a^3  \ddot{a} +96 r^4 \dot{a}^4
-96 r^4 a \dot{a}^2 \ddot{a}
\Biggl[
2 r f''+f' \left(3 r \gamma '+4\right)+f \Bigl(2 r \gamma ''+\gamma ' \left(r \gamma '+4\right)\Bigr)\ddot{a}
\Biggr] \nonumber\\
&&+
a^4 f^2 e^{2 \gamma}\Biggl[
16 \left(r^2 f'^2+1\right)+r^4 \left(2 f''+3 f' \gamma '\right)^2
+f^2 \Bigl(4 r^4 \gamma ''^2+4 r^4 \gamma '^2 \gamma ''+\left(r^2 \gamma '^2+4\right)^2\Bigr)\nonumber\\
&& +2 f \Biggl(2 r^4 f'' \left(2 \gamma ''+\gamma '^2\right)+r^2 f' \gamma ' \Bigl(3 r^2 \left(2 \gamma ''+\gamma '^2\right)+8\Bigr)-16\Biggr)
\Biggr]+8 r^2 a^2\Biggl[6 r^2 \ddot{a}^2-e^{\gamma} \dot{a}^2\nonumber\\
&&\times\Biggl(
2 r^2 f'^2+f \Bigl[r^2 \left(f' \gamma '-2 f''\right)+f \left(\left(r \gamma '-2\right)^2-2 r^2 \gamma ''\right)-4\Bigr]
\Biggr)
\Biggr]
\Biggr\},
    \label{eq:Kretschmann}
\end{eqnarray}
\end{widetext}
in which overdots and primes correspond, respectively, to differentiations with respect to conformal time and the radial coordinate. With the condition \eqref{10}, the Kretschmann scalar given above is reduced to 
\begin{widetext}
\begin{equation}\label{Kreduced}
    \mathcal{K}=\frac{2}{a}\left[12\dot{a}^4-12a\dot{a}^2\ddot{a}+\frac{a^4}{r^4}\left(2(-1+f)^2+r^2f^{\prime\;2}\right)+2a^2\left(3\ddot{a}^2-\frac{2\dot{a}^2(-1+f+rf^{\prime\;2})}{r}\right)\right],
\end{equation}
\end{widetext}
which, in the cosmological limit, recovers the one given in the FLRW spacetime.

\section{Separated results of data analysis}
\label{sec:separated}
In this appendix, we show the 1D and 2D posterior distributions separately to visualize more clearly the confidence regions obtained for the parameters of the model in each case for different datasets. 
\onecolumngrid
\begin{center}
\begin{figure}[h!]
  \includegraphics[width=0.55\textwidth]{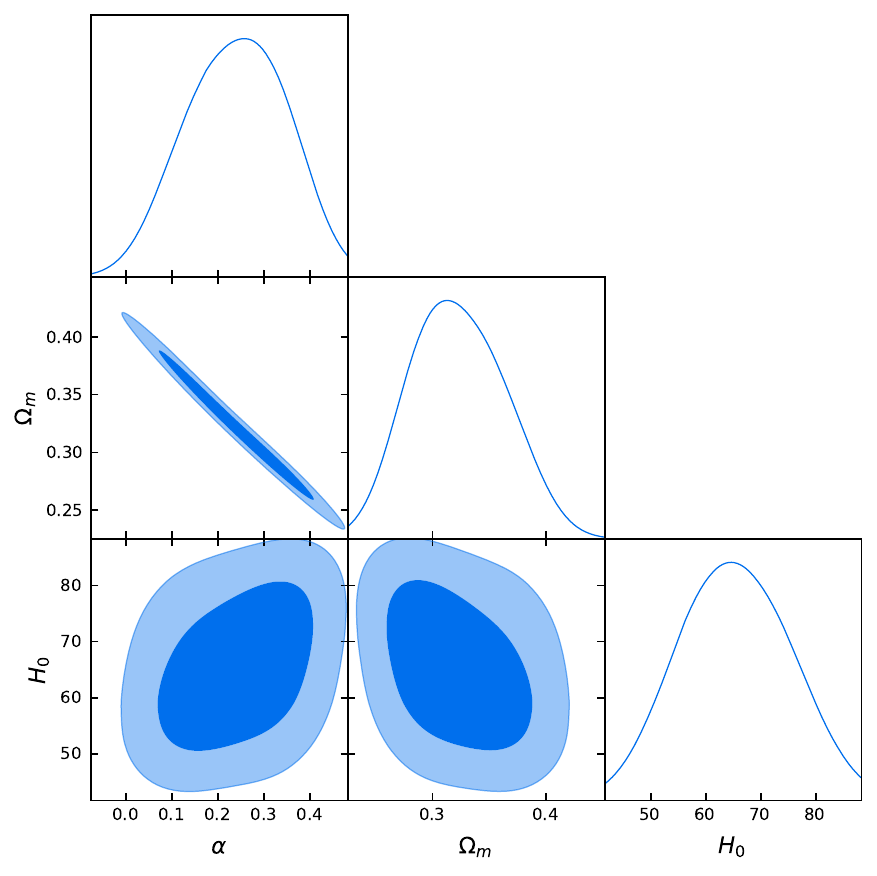}
\caption{1D and 2D marginalized posterior distributions for the parameters $\Omega_m$, $H_0$ and $\alpha$ using only data from SDSS DR12 (BAO).}
\label{fig:BAO}
\end{figure}

\begin{figure}[h!]
  \includegraphics[width=0.55\textwidth]{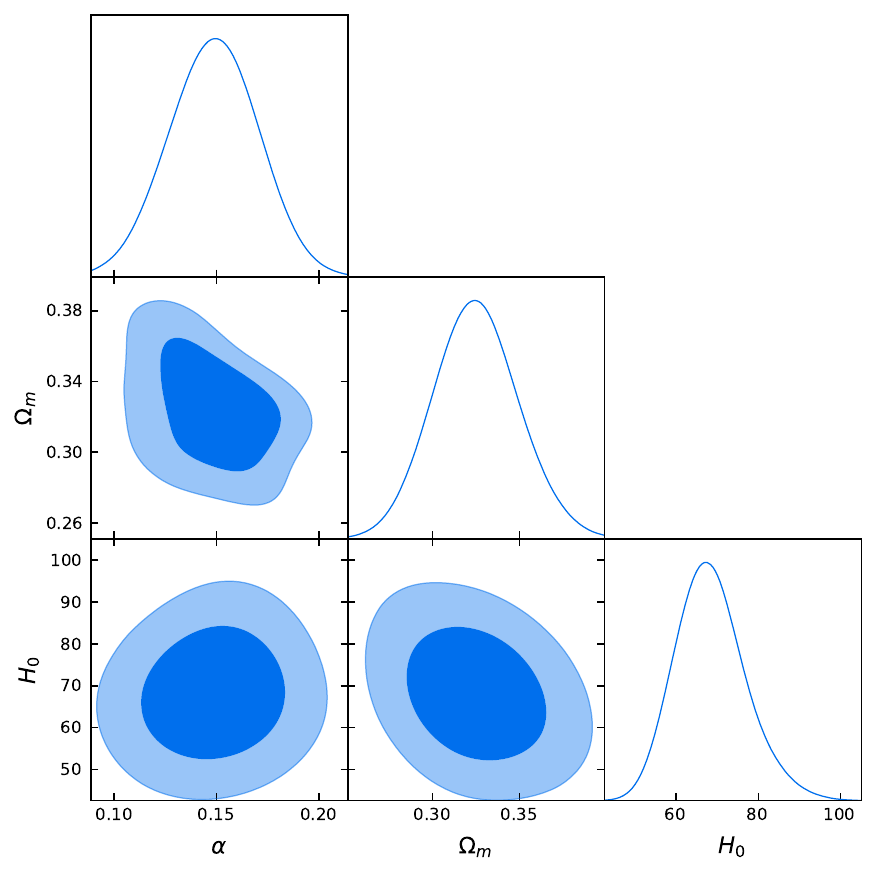}
\caption{1D and 2D marginalized posterior distributions for the parameters $\Omega_m$, $H_0$ and $\alpha$ using only data from Union 2.2 (SN Ia).}
\label{fig:UNION}
\end{figure}

\begin{figure}[h!]
  \includegraphics[width=0.55\textwidth]{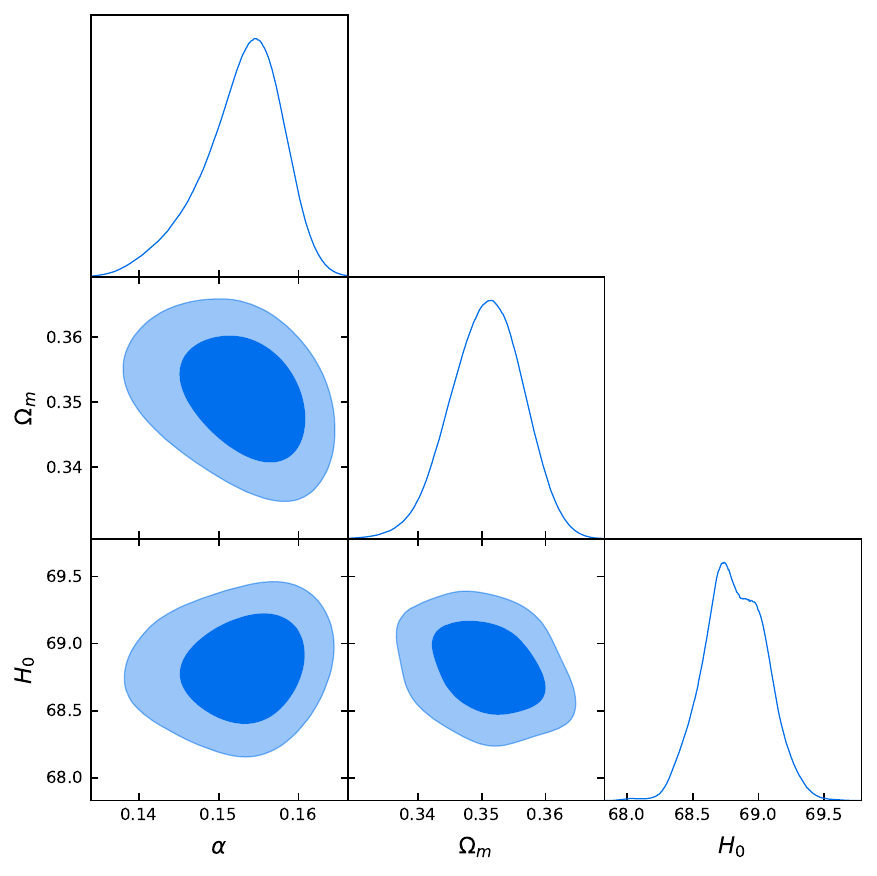}
\caption{1D and 2D marginalized posterior distributions for the parameters $\Omega_m$, $H_0$ and $\alpha$ using SDSS DR 12 $+$ Union 2.2.}
\label{fig:BAOUNION}
\end{figure}
\end{center}
\twocolumngrid

\bibliographystyle{ieeetr}
\bibliography{oja_template}
\end{document}